\documentclass[prb,twocolumn,aps,amsmath, amssymb, english,showpacs,longbibliography,superscriptaddress]{revtex4-2}

\usepackage{natbib}
\usepackage{float}
\usepackage{graphicx}
\usepackage{dcolumn}
\usepackage{tabularx}
\usepackage{bm}
\usepackage[dvipsnames]{xcolor}
\usepackage[normalem]{ulem}

\begin{document}

\preprint{APS/123-QED}

\title{Modeling the diversity of laser-induced spin dynamics in Gd/FeCo multilayers}

\author{Aleksandr Buzdakov}
\altaffiliation{Current affiliation: Istituto Italiano di Tecnologia, Genoa, Italy}
\affiliation{Interactive Fully Electrical Vehicles Srl, Turin, Italy}
\author{Thomas Blank}
\affiliation{Radboud University, Institute for Molecules and Materials, 6525 AJ Nijmegen, the Netherlands}
\author{Konstantin Zvezdin}
\affiliation{Istituto P.M., Turin, Italy}
\author{Alexey Kimel}
\affiliation{Radboud University, Institute for Molecules and Materials, 6525 AJ Nijmegen, the Netherlands}
\author{Oksana Chubykalo-Fesenko}
\affiliation{Instituto de Ciencia de Materiales de Madrid, CSIC, Madrid, Spain}

\date{\today}

\begin{abstract}
Recent experimental findings revealed an exceptionally diverse laser-induced spin dynamics tunable by magnetic field ($H$) and temperature ($T$) in ferrimagnetic Gd/FeCo multilayers, however the theoretical picture of these processes remains unclear.
To bridge this gap, we theoretically explore $H-T$ phase diagram of such a ferrimagnet and perform a modeling of the laser-induced spin dynamics using the Landau-Lifshitz-Bloch equation.
Our model can describe both transverse and longitudinal spin dynamics in the ferrimagnetic multilayer, including ultrafast helicity-independent all-optical switching, observed experimentally.
We explore the magnetic $H-T$ phase diagram and the full range of magnetization switching at low laser fluences.
We also examine the exchange relaxation mechanism critical for ultrafast switching
at higher laser fluences.
Our theoretical findings closely match experimental results, demonstrating the validity of the proposed models and their ability to predict static and dynamic magnetic properties of ferrimagnetic multilayers as functions of magnetic field and temperature.
\end{abstract}

\maketitle

\section{Introduction}
Ferrimagnets are possibly the most appealing  systems in ultrafast magnetism.
The discovery of all-optical magnetic switching in GdFeCo amorphous alloy nearly two decades ago boosted the field of ultrafast magnetism and eventually led to discoveries of many counter-intuitive phenomena in rare-earth transition metal alloys \cite{radu_2011, ostler_2012, bokor_2017}, including toggle switching \cite{radu_2011, ostler_2012} and current-induced switching \cite{bokor_2017}.
It has been also realized that rare-earth/transition-metal multilayers offer additional degrees of freedom in the field \cite{beens_2019}.
Stacking layers of different chemical elements allows one to control the interlayer exchange interaction, magnetisation and magnetic anisotropy of the artificial ferrimagnet.
As a result, thicknesses of the multilayers can be tuned with the aim to control compensation and Curie temperatures such that a subtle tuning of the applied magnetic field or temperature can result in dramatic changes of laser-induced spin dynamics, which can acquire uniquely diverse forms. This work is inspired by recent experimental findings of exceptionally diverse laser-induced spin dynamics tuneable with the help of magnetic field ($H$) and temperature ($T$) in ferrimagnetic Gd/FeCo multilayers \cite{blank_2024}.

\begin{figure}[h!]
\includegraphics[width=1.0\linewidth]{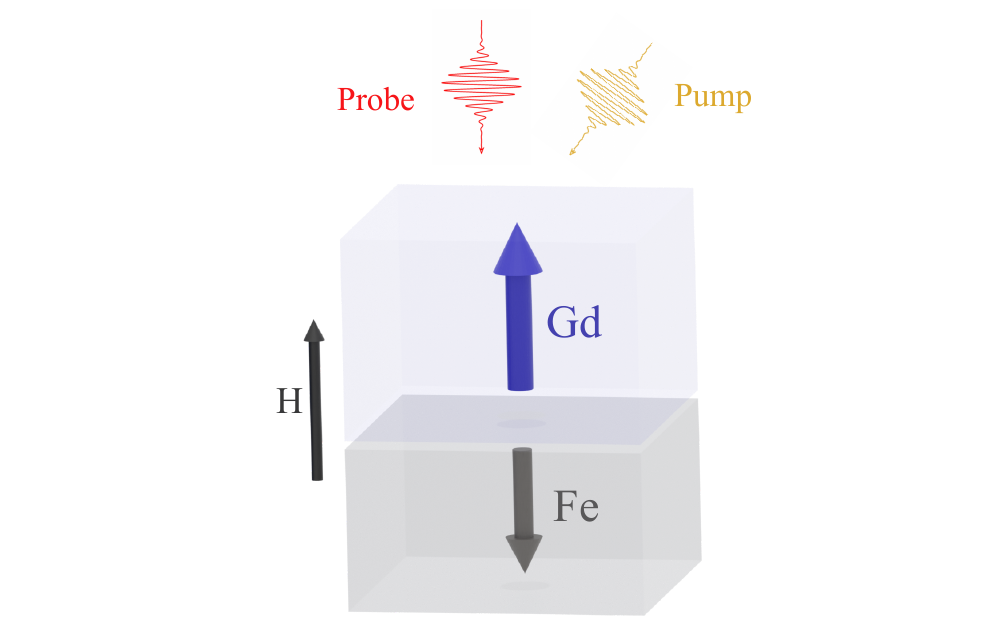}
\caption{\label{fig:structure} Sketch of the studied multilayered structure in two-sublattice mean-field approximation.}
\end{figure}
\par

Surprisingly, despite a significant progress made through diverse theoretical approaches –- ranging from atomistic modeling \cite{ostler_2012, jakobs_atxitia_2022}, the s-d model \cite{remy_2023} to the microscopic three-temperature model \cite{beens_2019} -- modeling which is able to reproduce such a diverse spin dynamics has not been reported so far.
Particularly, the majority of the previously published theoretical and computational studies of ultrafast magnetism in ferrimagnetic materials have been focused on systems with collinear magnetization direction.
At the same time, an application of moderate magnetic field to ferrimagnetic multilayers can significantly cant their local magnetization direction.
Earlier efforts to model the equilibrium and ultrafast magnetism of canted ferrimagnets using thermodynamic and Lagrangian principles were published in \cite{davydova_2019, davydova_zvezdin_2019, pogrebna_2019}, but these studies lacked a proper formulation for temperature-dependent longitudinal relaxation, which is crucial for modeling temperature-driven magnetization dynamics.
Here we propose an approach which fills that gap.

In particular,  we treat the system as two interconnected sublattices -- FeCo and Gd (see Fig.~\ref{fig:structure}). We use the mean-field approximation  for their temperature-dependent magnetisation  and we construct the static $H-T$ phase diagram by employing the self-consistent Curie-Weiss equations.
For modeling magnetisation dynamics we consider that each sublattice is governed by a macrospin described by the Landau-Lifshitz-Bloch (LLB) equation.
The approach to modeling ultrafast spin dynamics in ferrimagnets using the LLB equation was initially derived in \cite{atxitia_nieves_2012} and subsequently utilized in \cite{atxitia_2013, suarez_2015, blank_2022}. Being a high-temperature extension of the Landau-Lifshitz-Gilbert (LLG) equation, the LLB approach allows a simple interpretation of precessional dynamics but removes the constraint of conserved magnetisation length and includes  longitudinal magnetisation dynamics. Importantly, we generalize the original two-sublattice LLB by including the  exchange relaxation, allowing to accurately represent the longitudinal momentum transfer between sublattices as proposed in \cite{baryakhtar_2013, jakobs_atxitia_2022,jakobs_atxitia_2022a}. With the improved model  we are able to fill the modeling gap by reproducing static and dynamical behaviour of ferrimagnetic multilayer in a wide range of external magnetic fields and temperatures.

\par
The paper is organized as follows: Section II presents  the calculated $H-T$ phase diagrams.
In Sec. III we first describe the methods used to calculate magnetization dynamics.
The two subsequent subsection  focus on transverse and longitudinal magnetization dynamics, respectively.
Section IV provides conclusions and summarizes the open questions.

\section{Static $H-T$ phase diagrams.}
\par
\begin{figure}
\includegraphics[width=0.9\linewidth]{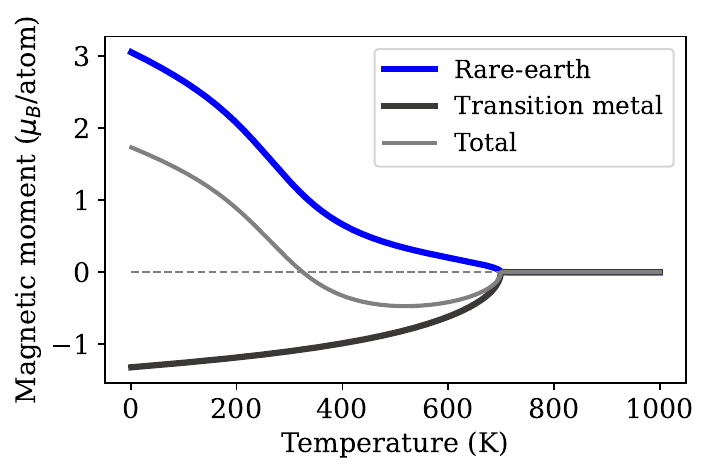}
\caption{\label{fig:m_vs_t_example} Magnetic moment versus temperature dependence calculated using the self-consistent Curie-Weiss equation (\ref{eqn:curie-weiss}) following the procedure explained in the text. Parameters were taken from Ref.~\cite{blank_2022}.}
\end{figure}
In order to simulate the static $H-T$ phase diagram of the ferrimagnetic multilayer, 
 we solve the self-consistent Curie-Weiss equations in the mean-field approximation (MFA):
\begin{equation}\label{eqn:curie-weiss}
\textbf{m}_{\nu} = L(\beta \mu_{\nu} H_{\mathrm{eff},\nu}) \frac{\textbf{H}_{\mathrm{eff},\nu}}{H_{\mathrm{eff},\nu}},
\end{equation}
where $\nu = $ FeCo, Gd. 

Effective fields acting on each sublattice have the following form: $\mu_{\nu} \textbf{H}_{\mathrm{eff},\nu} = q_{\nu} z J_{\nu} \textbf{m}_{\nu} + q_{\kappa} z J_{\nu-\kappa} \textbf{m}_{\kappa} + \mu_{\nu} \textbf{H}_{ext} + \mu_{\nu} \textbf{H}_{A,\nu}$, $\mu_{\nu} \textbf{H}_{A,\nu} = 2 K_{\nu}(\textbf{m}_{\nu} \cdot \textbf{z}) \textbf{z}$, where $\textbf{z}$ is the easy-axis of anisotropy, $J_{\nu}$ represents the intra-sublattice and $J_{\nu-\kappa}$ the inter-sublattice exchange parameter, $q_{\nu}$ is the  ion concentration, $q_{\mathrm{Gd}}=0.4$, $q_{\mathrm{FeCo}}=1-q_{\mathrm{Gd}}$ (i.e. the relative thickness of the layers in the multilayer case), $z=6$.
\par
The magnetic parameters derived from this study are summarized in the accompanying table \ref{table:parameters}. They are chosen with the aim to closely reproduce the experimental state diagram reported in Ref.~\cite{blank_2024}.
We should point out how remarkably exchange between (in our case much smaller) FeCo and Gd subsystems in out multilayered structure differs from the exchange typical for GdFeCo alloys.
The solution of the two self-consistent equations (\ref{eqn:curie-weiss}) is shown in Fig.~\ref{fig:m_vs_t_example}.
One can see how magnetic moment of each sublattice changes upon temperature increase.
Firstly, the magnetic moment of the rare-earth sublattice is larger than the magnetic moment of the transition metal sublattice.
Upon temperature increase the magnetic moment of each sublattice decreases, but at one point the magnetic moment of rare-earth sublattice becomes smaller that the magnetic moment of transition metal sublattice and the total magnetic moment re-orients in the opposite direction.
The magnetisation compensation point is around $T=320$ K. 
Eventually, upon further increase of temperature, the magnetic moment of each sublattice becomes zero.
\begin{figure}
\includegraphics[height=12cm]{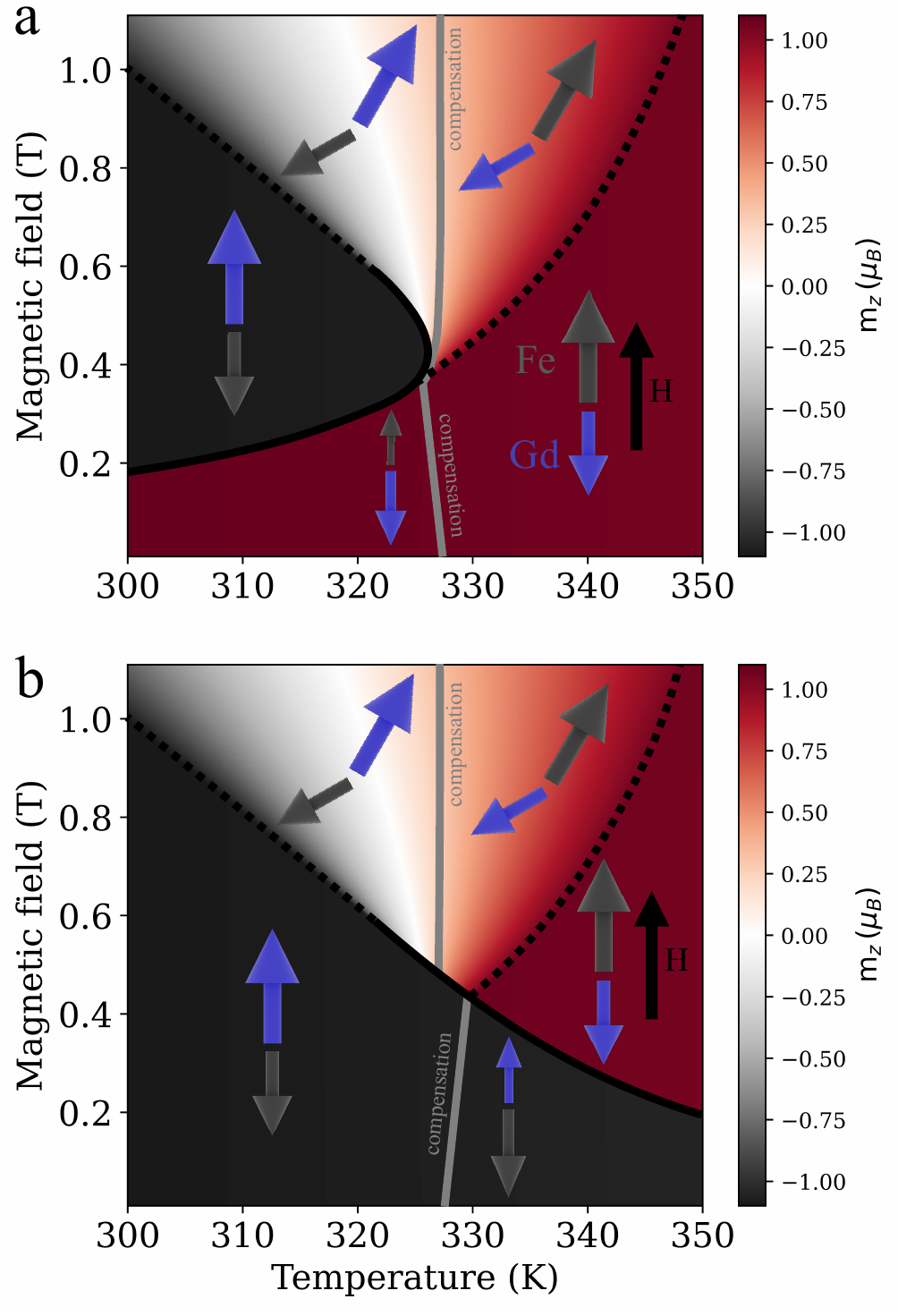}
\caption{\label{fig:static_phase_diag_Kfe_1e-24} Static $H-T$ phase diagrams for magnetization (normalized magnetization times $\mu_{\mathrm{FeCo}}$) of FeCo sublattice of Gd/FeCo multilayer calculated with additional easy-axis magnetic anisotropy of FeCo subsystem ($K_{\mathrm{FeCo}} = 10^{-24} \, J$, $K_{\mathrm{Gd}} = 0 \, J$). Panels (a) and (b) correspond to two distinct solutions of the self-consistent equation. The case (a) correspond to the situation when one starts at zero field with FeCo in $+z$ direction while the case (b) corresponds to the situation when at zero field Gd is perpendicular to the film. A dashed line indicates second-order phase-transition, solid lines indicates first-order phase-transition. Grey arrow corresponds to FeCo subsystem, blue arrow corresponds to Gd subsystem.}
\end{figure}
\begin{figure}
\includegraphics[height=12cm]{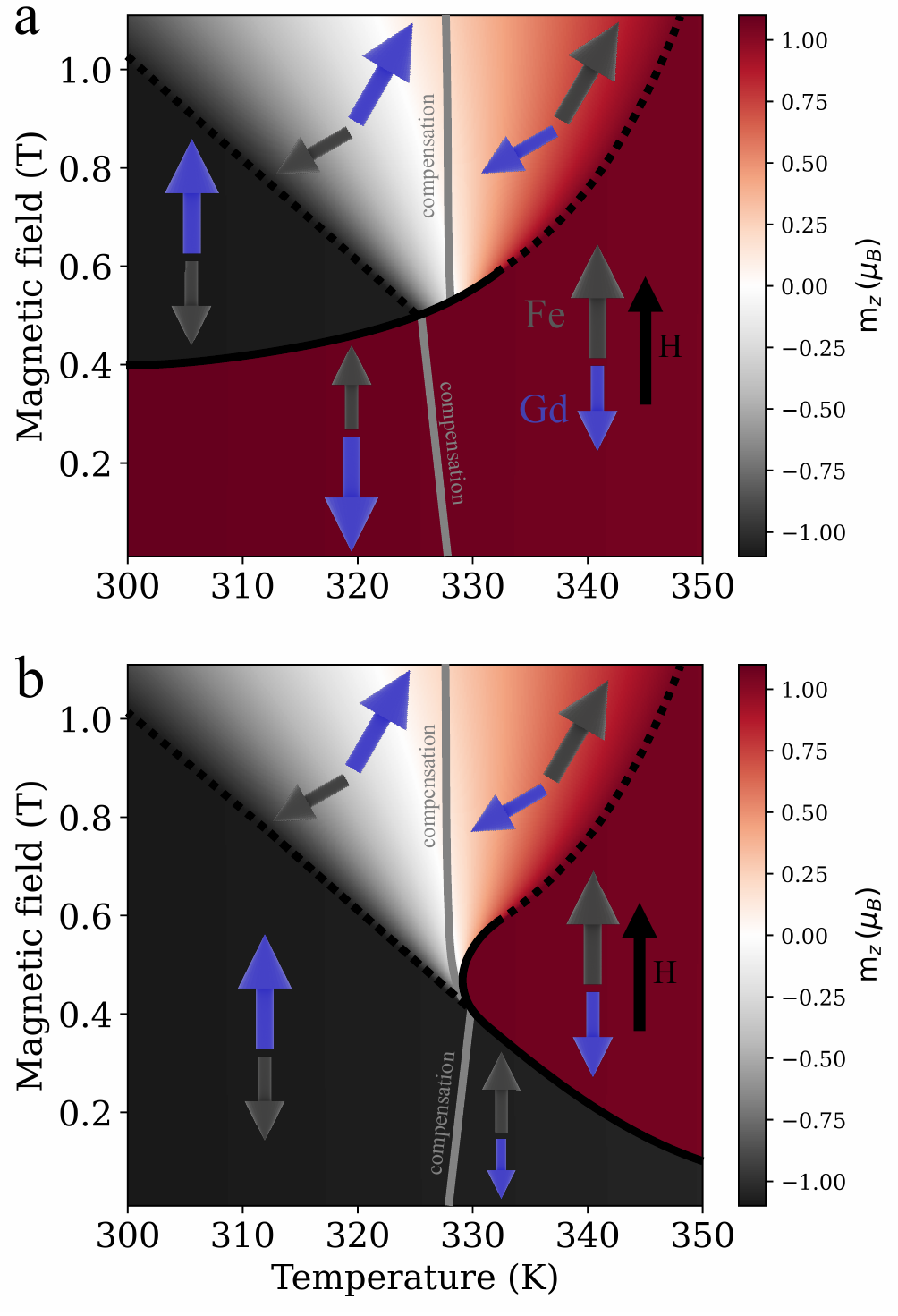}
\caption{\label{fig:static_phase_diag_Kgd_10e-24} Static $H-T$ phase diagrams for magnetization (normalized magnetization times $\mu_{\mathrm{FeCo}}$) of FeCo sublattice of Gd/FeCo multilayer calculated with additional easy-axis magnetic anisotropy of Gd subsystem ($K_{\mathrm{Fe}} = 0 \, J$, $K_{\mathrm{Gd}} = 10 \times 10^{-24} \, J$). Panels (a) and (b) correspond to two distinct solutions of the self-consistent equation. A dashed line indicates second-order phase-transition, solid line indicates first-order phase-transition. Grey arrow corresponds to FeCo subsystem, blue arrow corresponds to Gd subsystem.}
\end{figure}
\par
The self-consistent equations allow us to construct the complete $H-T$ phase diagrams.
External magnetic field in all of our results is oriented along the $z$-axis, which in turn is oriented perpendicular to the surface of the sample.
Figures~\ref{fig:static_phase_diag_Kfe_1e-24},\ref{fig:static_phase_diag_Kgd_10e-24} presents $H-T$ phase diagrams for various cases of easy-axis magnetic anisotropy within the system, where color indicates the $z$ component of the FeCo subsystem magnetization (normalized magnetization times $\mu_{\mathrm{FeCo}}$).
In the case of no magnetic anisotropy, every parameter set leads to a second-order, or continuous, phase-transition.
It means that the order parameter (e.g. magnetization) is characterized by a smooth change (i.e. reversible processes)  upon changing the external field or temperature.
\par
Introducing easy-axis magnetic anisotropy into the transition metal subsystem gives two distinct solutions of the self-consistent equation (\ref{eqn:curie-weiss}), given by the two distinct orientation of the antiferromagnetic vector as an initial condition, parallel and antiparallel to the external magnetic field (see Fig.~\ref{fig:static_phase_diag_Kfe_1e-24}).
The first (Fig.~\ref{fig:static_phase_diag_Kfe_1e-24}a) corresponds to the situation when we start at zero field with FeCo in $+z$ direction while the other one (Fig.~\ref{fig:static_phase_diag_Kfe_1e-24}b) corresponds to the situation when at zero field Gd is perpendicular to the film.
The one corresponding to the magnetisation of the dominant sublattice parallel to the field is stable while the one with the sublattice of a smaller magnetisation parallel to the external field is metastable.
Also, the added easy-axis magnetic anisotropy resulted in the appearance of first-order phase transitions in the system.
The first-order phase-transition is characterised by the appearance of discontinuity in the first-order derivative of the thermodynamics potential of the system (e.g. magnetization) and thus hysteresis when the external parameter is changed.
The first order phase transitions in Fig.~\ref{fig:static_phase_diag_Kfe_1e-24} are depicted by solid black line, while the smooth second order phase transitions are depicted by black dashed lines.
It is important to mention that in this case of uniaxial anisotropy the solid black line is inclined to the left side.
Also note the existence of the triple point in Fig.~\ref{fig:static_phase_diag_Kfe_1e-24} when the line of first-order transition concludes at a specific temperature and magnetic field strength, giving way to a line of second-order transitions.
In theory, a tricritical point on a $H-T$ phase diagram is common to magnets with antiferromagnetically coupled sublattices.
Hence, experimentally, it can also be found in many other ferri- and antiferromagnets.
\par
When anisotropy is introduced to, or is prevalent in, the rare-earth subsystem, the situation is similar as before, we observe the appearance of the first-order phase transition (see Fig.~\ref{fig:static_phase_diag_Kgd_10e-24}).
But in that case, the first-order transition line inclines to the right side and the triple point appears in the situation when at zero field we start with Gd perpendicular to the thin film plane.
The simulated phase diagram is consistent with the experimental phase diagram reported in \cite{blank_2024}, showing that the anisotropy of the rare-earth subsystem plays a dominant role in our system.
Such a behavior has previously been observed in the ferrimagnetic GdFeCo materials as detailed in \cite{davydova_2019, pogrebna_2019}.
\par

\begin{table}
\begin{center}
\begin{tabular}{ | c | c | }
 \hline
  Parameter & Value \\
 \hline
 $\mu_{\mathrm{FeCo}}$ & 2.217 $\mu_{B}$ \\
 \hline
 $\mu_{\mathrm{Gd}}$ & 7.63 $\mu_{B}$  \\
 \hline
 $J_{\mathrm{FeCo}}$ & $8 \times 10^{-21} \, \mathrm{J}$ \\
 \hline
 $J_{\mathrm{Gd}}$ & $5.95 \times 10^{-21} \, \mathrm{J}$ \\
 \hline
 $J_{\mathrm{FeCo-Gd}}$ & $- 0.05 \times 10^{-21} \, \mathrm{J}$ \\
 \hline
 $K_{\mathrm{FeCo}}$ & $ 0 $ \\
 \hline
 $K_{\mathrm{Gd}}$ & $10 \times 10^{-24} \, \mathrm{J}$ \\
 \hline
\end{tabular}
\caption{\label{table:parameters} Parameters used in the model.}
\end{center}
\end{table}

\section{Modeling the diversity of ultrafast magnetisation dynamics}
\subsection{Methodology}

Having reproduced the main features of $H-T$ phase diagram experimentally deduced in Ref.~\cite{blank_2024}, it is natural to ask if one can further develop the model and reproduce the diversity of laser-induced spin dynamics, as also experimentally observed in Ref.~\cite{blank_2024}. 
We make use of the well-known two-sublattice classical LLB equation, which accounts not only for transverse, but also longitudinal relaxation \cite{garanin_1997}:
\begin{equation}
\begin{split}
\dot{\textbf{m}}_{\nu} = -\gamma_{\nu} \left[ \textbf{m}_{\nu} \times \textbf{H}_{\mathrm{eff},\nu} \right] - \Gamma_{\nu,\parallel} \left( 1 - \frac{\textbf{m}_{\nu}\textbf{m}_{0,\nu}}{m_{\nu}^{2}} \right) \textbf{m}_{\nu} \\ - \Gamma_{\nu,\perp} \frac{[\textbf{m}_{\nu}\times[\textbf{m}_{\nu}\times\textbf{m}_{0,\nu}]]}{m_{\nu}^{2}}.
\label{eq:LLBpar}
\end{split}
\end{equation}
where $\nu$ stands either for RE or TM sublattice. Here the transverse relaxation rate is written as follows:
\begin{equation}
\Gamma_{\nu, \perp} = \frac{\Lambda_{\nu, N}}{2} \left[ \frac{\xi_{0,\nu}}{L(\xi_{0,\nu})} - 1 \right].
\end{equation}
and the longitudinal relaxation rate is:
\begin{equation}
\Gamma_{\nu, \parallel} = \Lambda_{\nu, N} \frac{L(\xi_{0,\nu})}{\xi_{0,\nu} L^{'}(\xi_{0,\nu})},
\end{equation}
where the N\'eel attempt frequency is $\Lambda_{\nu, N} = \frac{2 \gamma_{\nu} \lambda_{\nu}}{\beta \mu_{\nu}}$, the normalized magnetization in equilibrium is $\textbf{m}_{0,\nu}=L(\xi_{0,\nu})\frac{\boldsymbol\xi_{0,\nu}}{\xi_{0,\nu}}$, the so-called reduced effective field is $\boldsymbol\xi_{0,\nu}=\beta \mu_{\nu} \textbf{H}_{\mathrm{eff},\nu}$, $\beta = 1/k_{B}T$, $T$ is the temperature, $k_B$ is the Boltzmann constant,  the Langevin function is $L(\xi) = \coth (\xi) - 1/\xi$ and the derivative of the Langevin function is $\frac{dL(\xi)}{d\xi} = 1/\xi^2 - 1/\sinh^2\xi$, $\xi_{0,\nu}=\left| \boldsymbol\xi_{0,\nu} \right|$ and $\lambda_{\nu}$ is the coupling-to-the bath parameter ("atomistic damping parameter").
Using the LLB equation, we can calculate normalized magnetization dynamics in the mean-field two-sublattice approximation.

The LLB Eq.(\ref{eq:LLBpar}) can describe  the transfer of the angular momentum from one sub-lattice to another via the transverse magnetization components \cite{atxitia_2013} but the detailed analysis shows that it cannot describe the switching with pure longitudinal dynamics. 

A more phenomenological approach which accounts for angular momentum exchange between the magnetic sublattices for pure longitudinal magnetization dynamics, is based on the theory of V. G. Baryakhtar \cite{baryakhtar_1998} and has been used in \cite{baryakhtar_2013,mentink_2012} to describe ultrafast magnetisation dynamics in RE-TM alloys.
This approach is relying on Onsager's reciprocal relations and uses the dissipative function $Q$ for a two-sublattice ferrimagnet (with sublattices $A$ and $B$) in the following form:
\begin{equation}
Q = \lambda(\textbf{H}_{A} - \textbf{H}_{B})^2.
\end{equation}
Meaning that relaxation term $R_{i}$ for $i$-th sublattice will take the following form:
\begin{equation}\label{eqn:r_func}
\textbf{R}_{i} = \delta Q / \delta \textbf{H}_{i} = \lambda (\textbf{H}_{A} - \textbf{H}_{B}).
\end{equation}
Here $\textbf{H}_{A,z} = -\frac{\delta \mathrm{w}}{\delta \textbf{m}_{A,z}}$ and $\textbf{H}_{B,z} = -\frac{\delta \mathrm{w}}{\delta \textbf{m}_{B,z}}$ are
effective fields acting on the sublattices $A$ and $B$, $\lambda$ is the coupling to the bath parameter and  $\mathrm{w}$ is the reduced free energy. In the simplest case  we have $\mathrm{w} = \frac{J_{A} \textbf{m}^2_{A}}{2} + \frac{J_{B} \textbf{m}^2_{B}}{2} + \frac{J_{AB}}{2} \textbf{m}_{A} \textbf{m}_{B}$ and for the relaxation term we get:
\begin{equation}
R_{A,z} \propto \lambda J_{AB} (m_{B,z} - m_{A,z})
\end{equation}

Here we want to show an alternative view on that problem. Indeed, the correct LLB equation should be consistent with the formulation of the free energy and should have the form $\dot{\mathbf{m}}=-\Lambda_1 \frac{\partial \mathcal{F}}{\partial \mathbf{m} } + \Lambda_2 [\mathbf{m} \times  \frac{\partial \mathcal{F}}{\partial \mathbf{m} }] $, where $\Lambda_1$ and $\Lambda_2$ are longitudinal and transverse relaxation parameters. We will make use of the free energy for two-sublattice magnet obtained with the help of variational procedure in \cite{pablo_phdthesis}. The main problem noticed in the work \cite{pablo_phdthesis} is that the standard MFA derivation also used in the two-sublattice LLB equation gives an incorrect dynamical behavior near the critical temperature and should be corrected.
For the pure longitudinal case, free energy per spin in the physically correct derivation is given by:
\begin{equation}
\begin{split}
\mathcal{F} = &- \frac{1}{\beta} \ln{(4\pi)} - \frac{x_{A}}{\beta} \Lambda(\xi_{A,z}) - \frac{x_{B}}{\beta} \Lambda(\xi_{B,z})  \\ &- \frac{x_{A}}{2} J_{0,A} L^2(\xi_{A,z}) - \frac{x_{B}}{2} J_{0,B} L^2(\xi_{B,z}) \\ &- x_{A} J_{0,AB} L(\xi_{A,z}) L(\xi_{B,z}) \\ &+ \frac{x_{A}}{\beta} \xi_{A,z} L(\xi_{A,z}) + \frac{x_{B}}{\beta} \xi_{B,z} L(\xi_{B,z})
\end{split}
\end{equation}
where $\xi_{A,z} = \beta [J_{A} m_{A,z} + J_{AB} m_{B,z}]$, $\Lambda(x) = \ln [\sinh(x)/x]$, $L(x) = \coth(x) - 1/x$.
If we now expand free energy close to the critical temperature, we  get and expression similar to the one used above, i.e.:
\begin{equation}
\begin{split}
\mathcal{F} \approx &- \frac{1}{\beta} \ln(4\pi) + D_{A^2} m_{A,z}^2 + D_{B^2} m_{B,z}^2 \\ &+ D_{AB} m_{A,z} m_{B,z} + \mathcal{O}(m^4)
\end{split}
\end{equation}
\begin{equation}
D_{\nu^2} = \frac{\beta}{18} x_{\nu} \left[ (3 - \beta J_{\nu}) J_{\nu}^2 + (3 - \beta J_{\kappa}) J_{\nu \kappa}^2 - 2 \beta J_{\nu \kappa}^2 J_{\nu} \right]
\end{equation}
\begin{equation}
\begin{split}
D_{\nu \kappa} = \frac{\beta}{9} x_{\nu} J_{\nu \kappa} [ (3 - \beta J_{\nu}) J_{\nu} + (3 - \beta J_{\kappa}) J_{\kappa} \\ - \beta (J_{\kappa} J_{\nu} + J_{\nu \kappa}^2) ]
\end{split}
\end{equation}
The intra-sublattice contributions are already correctly included in the longitudinal term of the original LLB equation so that we concentrate into the inter-sublattice ones.
Then, the corresponding relaxation term is given by:
\begin{equation}
R_{A,z} \propto - \tilde{\Lambda}_{1} \partial_{m_{A,z}} \mathcal{F} \propto - \tilde{\Lambda}_{1} \left[ \frac{\beta}{3} J_{AB}^2 m_{A,z} - \frac{\beta^2}{9} J_{AB}^3 m_{B,z} \right]
\end{equation}
Here  $\tilde{\Lambda}_{1}$ accounts to the inter-sublattice part of the relaxation only.
The critical temperature is given by $1/\beta = k_{B} T_{c} = J_{AB}/3$, so we finally get the expression for the relaxation term:
\begin{equation}
R_{A,z} \propto \tilde{\Lambda}_{1} J_{AB} (m_{B,z} - m_{A,z})
\end{equation}
Now we can see that we got the same relaxation term as from the approach based of Onsager's reciprocal relations.

Thus, in order to account for additional relaxation term given by Eq.(\ref{eqn:r_func}), one has to add additional term to the LLB equation. For the pure longitudinal relaxation it now reads::
\begin{equation}
\dot{m}_{A,z} = \gamma \alpha_{A} H_{A} + \gamma \frac{\alpha_{ex}}{\mu_{A}} (\mu_{A} H_{A} - \mu_{B} H_{B}).
\end{equation}
Here the first term corresponds to the longitudinal intra-sublattice relaxation term (as accounted in the original LLB equation Eq.(\ref{eq:LLBpar}) which can be cast in the form $\dot{m}_{A,z}= \Gamma_{\parallel,a} (m_{A,z}-m_{A,z0}) = \gamma \alpha_{A} H_A$.
The specific form and the value of the coefficient $\alpha_{ex}$, referred to as the exchange relaxation term,   will depend on the intrinsic scattering mechanisms. For example, in Ref.\cite{garanin_2009} it was shown to be defined by the spin-wave spectrum. Generally, other mechanisms can contribute and since the exchange relaxation acts on femtosecond timescale, one cannot expect this term even  to have a simple relation to the transverse relaxation.
Here the specific form of the  exchange relaxation coefficient was adopted from the works of Jakobs and Atxitia \cite{jakobs_atxitia_2022, jakobs_atxitia_2022a}, also related to the expansion of the exchange interactions:
\begin{equation}
\alpha_{ex} = \frac{1}{2} \left( \frac{\alpha_{A}}{z m_{A}} + \frac{\alpha_{B}}{z m_{B}} \right).
\end{equation}
\par
To introduce a laser pulse into the simulations, we make use of the so-called two-temperature model (2TM).
In particular, we assume that the laser pulse is fully absorbed by free electrons and the absorption increase the electron temperature $T_{e}$.
As a result of heat exchange between the electrons and the lattice, the electron $T_{e}$ and the lattice $T_{ph}$ temperatures equilibrate and, this process can be described with the following set of equations:
\begin{equation}
\begin{cases}
\frac{d T_{e}}{d t} = -\frac{G_{el-ph}}{\gamma_{e} T_{e}}(T_{e}-T_{ph}) + \frac{P(t)}{\gamma_{e} T_{e}}\\
\frac{d T_{ph}}{d t} = -\frac{G_{el-ph}}{C_{l}}(T_{ph}-T_{e})
\end{cases}
\end{equation}
where $C_{e} = \gamma_{e} T_{e}$ is the electron heat capacity, $C_{l}$ is the phonon heat capacity, $G_{el-ph}$ is the rate constant.
Here we assume that the laser pulse $P(t)$ has the Gaussian shape:
\begin{equation}
P(t) = P_{0} \exp (-4\ln2 (t-t_{0})^2/\tau^2),
\end{equation}
where the peak power is $P_0 \approx F/2\tau d$, $d$ is the sample thickness, $\tau$ is the pulse duration, $t_{0}$ is the time delay of the pulse.

\subsection{ Simulated  longitudinal magnetization dynamics}
\par
\begin{figure}
\includegraphics[height=16cm]{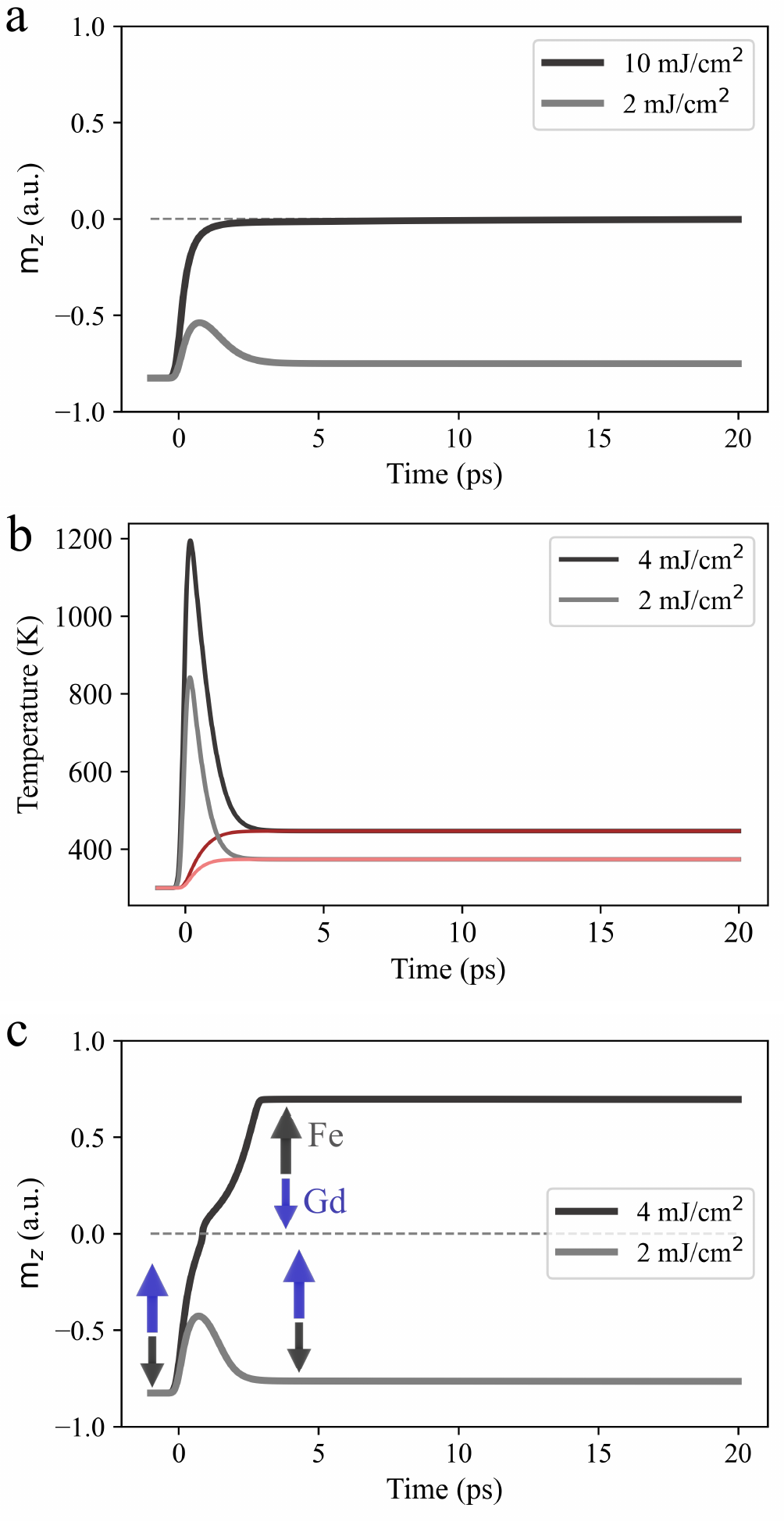}
\caption{\label{fig:fast_dynamics} Longitudinal magnetization dynamics for transition metal subsystem for the case $\mu_{0} H=0 \, \mathrm{T}$ and $T=300 \, \mathrm{K}$. Panel (a) shows the results of calculations using the non-modified ferrimagnetic Landau-Lifshitz-Bloch. No switching is observed upon an increase of the laser fluence. Panel (b) show the dynamics of the electron and the lattice temperatures. Panel (c) shows the results of the simulations with modified Landau-Lifshitz-Bloch equations. The simulations are performed for two fluences $2$ and $4 \, \mathrm{mJ}/\mathrm{cm}^2$, respectively.}
\end{figure}

We start with simulating laser-induced magnetisation dynamics at zero field. Here one expects pure longitudinal demagnetisation and recovery at small fluences while at large fluences longitudinal switching should be observed.
As we indicated above, in order to model large amplitude longitudinal magnetization dynamics and longitudinal (linear) switching, one has to account for an exchange of angular momentum between the magnetic sublattices during the longitudinal dynamics. While the original two-sublattice LLB equation correctly describes the degree of ultrafast demagnetisation, an increase of the laser fluence only leads to an increase of the degree of ultrafast demagnetization. This is illustrated in 
Figure \ref{fig:fast_dynamics} which  shows an example of the simulated transients from laser fluences $2 \, \mathrm{mJ}/\mathrm{cm}^2$ and $10 \, \mathrm{mJ}/\mathrm{cm}^2$, respectively.
\par
\begin{figure}
\includegraphics[height=12cm]{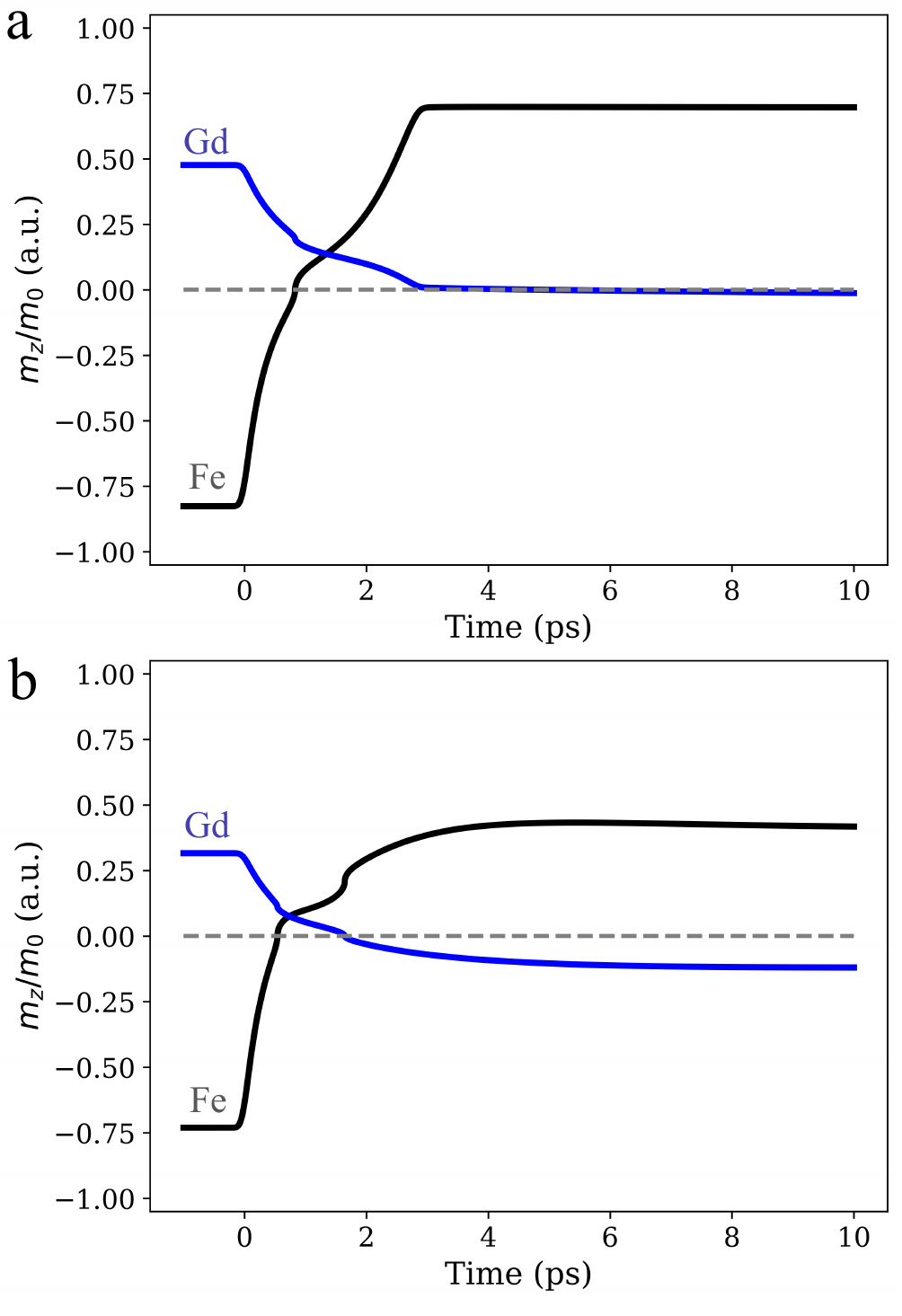}
\caption{\label{fig:gd_dynamics} Ultrafast magnetization dynamics in a ferrimagnet calculated for the case $\mu_{0} H=0 \, \mathrm{T}$ and $T=300 \, \mathrm{K}$ and $F = 4 \, \mathrm{mJ}/\mathrm{cm}^2$ using the magnetic parameters of the Gd/FeCo multilayer from Table 1 (panel (a)) and those of GdFeCo alloy (panel(b)). The parameters used for the simulations in the alloy are given in the text.}
\end{figure}

\par
\begin{figure}
\includegraphics[height=10cm]{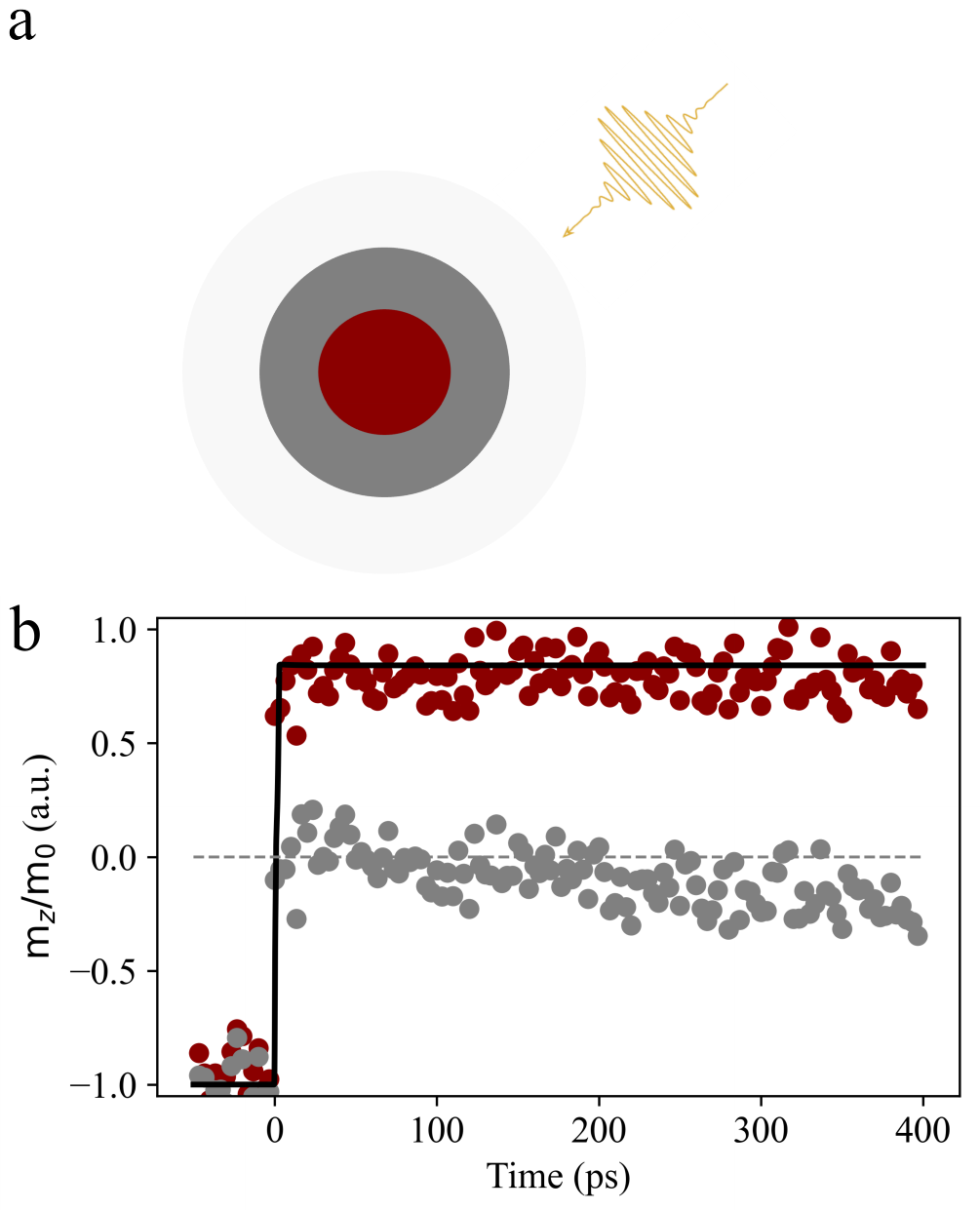}
\caption{\label{fig:fast_dynamics_expr} Ultrafast magnetization dynamics in Gd/FeCo multilayers at different pump fluences, $\mu_{0} H=0 \, \mathrm{T}$ and $T=300 \, \mathrm{K}$. Experimentally the fluence was varied across the laser beam having Gaussian spatial intensity distribution. Red circles are experimental data observed at $F \approx 8.8 \, \mathrm{mJ}/\mathrm{cm}^2$ in the area where most of the laser energy is concentrated. Grey circles correspond to the experimental data observed on the outer ring where much less energy is concentrated. The solid line show the results of the simulations. The dashed line indicates zero magnetization.}
\end{figure}

Adding the exchange relaxation term into the LLB equations and performing the simulations for a low laser pulse fluence ($F=2 \, \mathrm{mJ}/\mathrm{cm}^2$), we get once again ultrafast demagnetization (see grey line in Fig.~\ref{fig:fast_dynamics}c), while for a higher pulse fluence the longitudinal magnetization dynamics does not stop at the totally demagnetized state, but proceeds further towards magnetic switching (see black line in Fig.~\ref{fig:fast_dynamics}c).
One can see that the simulated dynamics has small discontinuities in its time-derivative (see Fig.~\ref{fig:gd_dynamics}).
The discontinuities correspond to the moments in time when the magnetizations approach zero and correspond to the divergence of the adopted unknown form of the exchange relaxation term from Refs.\cite{jakobs_atxitia_2022, jakobs_atxitia_2022a}.

Another interesting insight into the laser-induced longitudinal magnetization dynamics of ferrimagnetic multilayers can be obtained by comparison of the magnetization transients simulated for the parameters of the Gd/FeCo multilayers and for the parameters of a GdFeCo alloy ($J_{\mathrm{FeCo}} = 4.5 \times 10^{-21} \, \mathrm{J}$, $J_{\mathrm{Gd}} = 1.26 \times 10^{-21} \, \mathrm{J}$, $J_{\mathrm{FeCo-Gd}} = -1.09 \times 10^{-21} \, \mathrm{J}$, $q_{\mathrm{FeCo}} = 0.75$, $q_{\mathrm{Gd}} = 0.25$), respectively.
Figure \ref{fig:gd_dynamics} shows the corresponding dynamics and reveals that the behaviour of the rare-earth sublattice in the multilayer and in the alloy are drastically different. One can notice a complete demagnetisation of Gd in the multilayer case (Figure \ref{fig:gd_dynamics}a) while in the alloy case (Figure \ref{fig:gd_dynamics}b) Gd is polarized by FeCo and its magnetisation stays constant.
Thus, the exchange interaction between the spins of the rare-earth and transition metals play in the dynamics a decisive role.
\par
It is interesting to compare the outcome of the simulations of the longitudinal laser-induced magnetization dynamics in ferrimagnetic multilayers to the experimentally observed transients.
Figure~\ref{fig:fast_dynamics_expr} shows the experimentally observed normalized magnetization dynamics detected as described in Ref.~\cite{blank_2022} at $\mu_{0} H = 0 \, \mathrm{T}$ and $T = 300 \, \mathrm{K}$ for two laser fluences.
The laser fluence was changing across the laser beam in accordance with the Gaussian function.
Hence by detecting the laser-induced dynamics at a different point, one can reveal the dynamics at a different laser fluence.
It is seen that while one fluence results in a total ultrafast demagnetization, an increase of the fluence promotes ultrafast linear switching.
Our simulations c nicely reproduce this experimentally observed behavior (see the solid lines in Fig.~\ref{fig:fast_dynamics_expr}).

\subsection{Simulated transverse magnetisation dynamics}
\begin{figure}
\includegraphics[height=14cm]{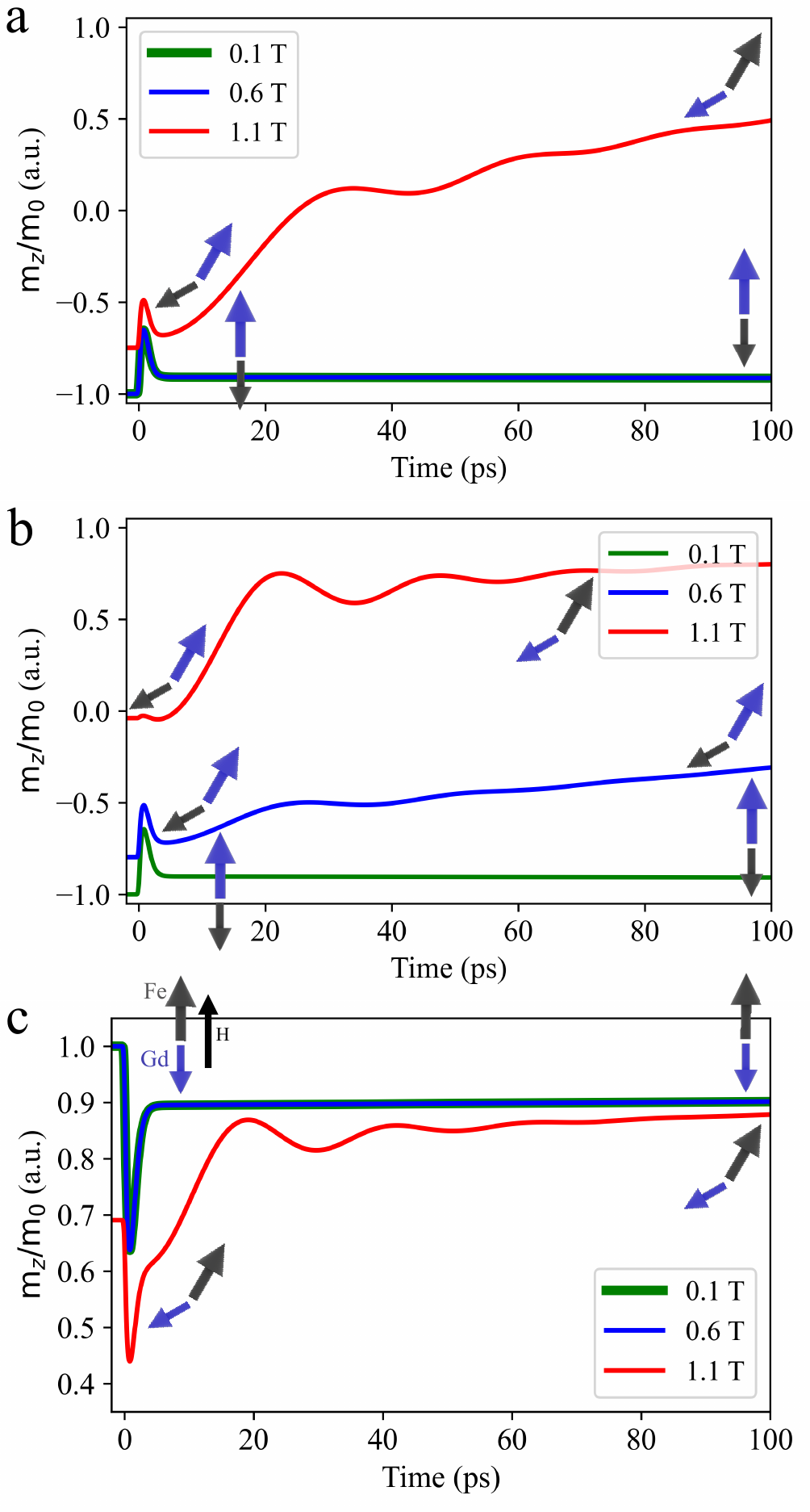}
\caption{\label{fig:slow_dynamics} Transverse magnetization dynamics of FeCo subsystem calculated using the model based on the Landau-Lifshitz-Bloch equations for moderate values of the pulses' fluence with different initial temperatures: a -- $300 \, \mathrm{K}$, b -- $320 \, \mathrm{K}$, c -- $340 \, \mathrm{K}$. Magnetic parameters of the Gd/FeCo multilayer are from Table 1.}
\end{figure}

We now turn our attention to the magnetisation dynamics in applied field when the predominant role of the transverse dynamics should be expected. Firstly, we set the initial temperature to $300 \, \mathrm{K}$, then we applied an external field, again equilibrate, and finally increase temperature using the 2TM.
The field is applied parallel with respect to the equilibrium orientation of the spins at $\mu_{0} H = 0 \, \mathrm{T}$.
The results are shown in Fig.~\ref{fig:slow_dynamics} in the Gd and FeCo dominated regions and for three different temperatures. 

Fig.~\ref{fig:slow_dynamics}a corresponds to Gd-dominated region before the magnetisation compensation point.  For low magnetic fields ($0.1 \, \mathrm{T}$, $0.6 \, \mathrm{T}$) we see only ultrafast (sub-ps) demagnetization and a partial recovery, i.e. the dynamics is similar to that observed in metallic ferromagnets \cite{koopmans_2009}.
For higher field ($1.1 \, \mathrm{T}$) we observe ultrafast demagnetization followed by magnetization switching (when the $z$-component $m_{z}$ of the normalized magnetization of the transition metal sublattice changes its sign).
The dynamics also reveals low amplitude and heavily damped oscillations.
For the higher initial temperature $T = 320 \, \mathrm{K}$ at low magnetic field ($0.1 \, \mathrm{T}$) we observe a similar ultrafast demagnetization (see Fig.~\ref{fig:slow_dynamics}b), while an increase of the field up to $0.6 \, \mathrm{T}$ changes the dynamics.
The demagnetization is followed by switching and oscillations. At $340 \, \mathrm{K}$ the dominating sublattice is FeCo.  The observed transients are qualitatively similar to those observed at $T = 300 \, \mathrm{K}$.
Comparing the dynamics with the phase diagram shown in Fig.~\ref{fig:static_phase_diag_Kgd_10e-24}, reveals that the switching and the oscillations are observed only if the laser pulse excites the ferrimagnet in a non-collinear state i.e. when the magnetizations of the two sublattices are no longer parallel.

\par
The observed magnetisation oscillations can be qualitatively explained using a simple model of two non-dissipative, but coupled Landau-Lifshitz equations corresponding to the magnetizations of the two magnetic sublattices.
\begin{equation}
\begin{cases}
\frac{1}{\gamma} \frac{\partial \textbf{m}_{A}}{\partial t} = [\textbf{m}_{A},\textbf{H}_{A}] \\
\frac{1}{\gamma} \frac{\partial \textbf{m}_{B}}{\partial t} =  [\textbf{m}_{B},\textbf{H}_{B}],
\end{cases}
\end{equation}
where the effective fields experienced by sublattices are $\textbf{H}_{i} = -\frac{\delta w}{\delta \textbf{m}_{i}}$ and systems' free energy is defined as follows $w = f_{A}(m_{A}^2) + f_{B}(m_{B}^2) + \frac{J_{AB}}{2} \textbf{m}_{A} \textbf{m}_{B}$.
Introducing the net magnetization $\textbf{m} \equiv \textbf{m}_{A} + \textbf{m}_{B}$ and antiferromagnetic vector $\textbf{l} \equiv l \hat{\textbf{l}} \equiv \textbf{m}_{A} - \textbf{m}_{B}$, our set of equations will transform into:
\begin{equation}
\frac{\partial \textbf{m}}{\partial t} = 0
\end{equation}
\begin{equation}
\begin{cases}
\frac{\partial l}{\partial t} = 0 \\
\frac{\partial \hat{\textbf{l}}}{\partial t} = \gamma J_{AB}  [\textbf{m},\hat{\textbf{l}}]
\end{cases}
\end{equation}

It is seen that since the Landau-Lifshitz equations do not contain any damping term, the net magnetization cannot change, while the antiferromagnetic vector $\textbf{l}$ can oscillate around $\textbf{m}$ with the characteristic frequency $\omega = \gamma J_{AB}$.

\section{Conclusions}
Here we theoretically explored laser-induced magnetization dynamics in a ferrimagnet as a function of applied magnetic field and temperature.
The goal of the study was to reproduce and to understand recently reported diverse laser-induced spin dynamics in Gd/FeCo multilayers.  We built a model allowing to simulate $H-T$ phase diagram and fitted the parameters with the aim  to reproduce the $H-T$ phase diagram observed experimentally.
Our results underline the role of a small inter-sublattice exchange parameter in the multilayer case responsible for a different magnetisation dynamics in comparison to the case of the alloy.
Particularly, small inter-sublattice exchange allows to experimentally access canted magnetic states of FeCo and Gd sublattices. In  this state, and especially close to magnetisation compensation point,  the magnetisation dynamics shows rich behavior as a function of applied field and temperature.

Using the parameters and modified Landau-Lifshitz-Bloch equations, we modeled both transverse and longitudinal magnetization dynamics and have been able to reproduce the experimentally observed dynamics in a broad range of magnetic fields and temperatures. The two-sublattice LLB equation has been modified based on the free-energy approach to account for longitudinal magnetisation switching casting it in the form consistent with previously used models \cite{baryakhtar_2013,mentink_2012,baryakhtar_1998,jakobs_atxitia_2022,jakobs_atxitia_2022a}.

We consider this work as an important step in understanding ultrafast magnetization dynamics with multisublattice materials with canted spin arrangements.

\section{Acknowledgements}
The project has received funding from the European Union’s Horizon 2020 research and innovation programme under the Marie Skłodowska-Curie Grant Agreement No.861300 (COMRAD) and European Research Council ERC Grant Agreement No.101054664 (SPARTACUS).

\bibliography{apssamp}

\end{document}